# Edge and bulk dissipative solitons in modulated $\mathcal{PT}$-symmetric waveguide arrays

YAROSLAV V. KARTASHOV[1,2,*] AND VICTOR A. VYSLOUKH[3]

[1]*ICFO-Institut de Ciencies Fotoniques, The Barcelona Institute of Science and Technology, 08860 Castelldefels (Barcelona), Spain*
[2]*Institute of Spectroscopy, Russian Academy of Sciences, Troitsk, Moscow, 108840, Russia*
[3]*Universidad de las Américas Puebla, Santa Catarina Mártir, 72820 Puebla, México*

**We address dissipative soliton formation in modulated $\mathcal{PT}$-symmetric continuous waveguide arrays composed from waveguides with amplifying and absorbing sections, whose density gradually increases (due to decreasing waveguide separation) either towards the center of the array or towards its edges. In such a structure the level of gain/loss at which $\mathcal{PT}$-symmetry gets broken depends on the direction of increase of the waveguide density. Breakup of the $\mathcal{PT}$-symmetry occurs when eigenvalues of modes localized in the region, where waveguide density is largest, collide and move into complex plane. In this regime of broken symmetry the inclusion of focusing Kerr-type nonlinearity of the material and weak two-photon absorption allows to arrest the growth of amplitude of amplified modes and may lead to the appearance of stable attractors either in the center or at the edge of the waveguide array, depending on the type of array modulation. Such solitons can be stable, they acquire specific triangular shapes and notably broaden with increase of gain/loss level. Our results illustrate how spatial array modulation that breaks $\mathcal{PT}$-symmetry "locally" can be used to control specific location of dissipative solitons forming in the array.**

More than one decade ago it was suggested that a class of parity-time ($\mathcal{PT}$) symmetric non-Hermitian Hamiltonians exists that admits completely real spectrum (see review [1] and references therein). This idea have penetrated from quantum mechanics to many other branches of physics, including optics. New degree of freedom connected with the possibility to engineer balanced gain/loss landscapes supporting stationary modes that propagate without net gain or attenuation, has opened broad prospects for shaping of light fields and design of optical structures with unusual operation regimes. Numerous optical $\mathcal{PT}$-symmetric systems were introduced, starting from directional couplers and coupled microresonators, to extended shallow photonic lattices and photonic crystals (see [2-4] for recent reviews).

The most representative feature of the $\mathcal{PT}$-symmetric system is the existence of the threshold level of gain/loss above which this symmetry gets broken and the spectrum becomes complex [1,5], the fact that was confirmed experimentally in [6,7]. Around this threshold many nonlinear and linear effects, such as power-controlled switching [8-10], soliton formation in periodic [11,12], localized [13,14], or truncated [15] structures, nonreciprocal soliton scattering [16], asymmetric mobility [17] and rectification [18], slow light [19], Anderson localization [20] and many others acquire nonconventional, unexpected features. $\mathcal{PT}$-symmetric lattices can be designed to support topological edge states [21]. Threshold for $\mathcal{PT}$-symmetry breaking depends on several factors, most notably on the size of the system and on the ratio of coupling constant between individual channels and amplitude of gain/loss. Usually, this threshold decreases with the increase in the number of elements in the system [22-24]. In multimode structures one may observe several "thresholds" defined by collisions of different pairs of modes, after which their eigenvalues move into complex plane and one of the modes start to grow, while other mode decays [25,26]. In nonlinear pseudopotentials an unusual regime is possible when $\mathcal{PT}$-symmetry cannot be broken at all [27]. Notice that $\mathcal{PT}$-symmetric solitons were constructed using uniform discrete lattices model [28], and exact analytical solutions for localized modes for solitons pinned to a parity-time-symmetric dipole were reported [29].

The aim of this Letter is to introduce finite $\mathcal{PT}$-symmetric system composed of multiple waveguides with varying separation (or density), where modulus of the coupling constant between waveguides and its ratio to gain/loss amplitude inside waveguides varies across the system. This variation results in localization of modes within particular domains, near the surface or in the center of the structure, and namely these localized modes lead to "local" $\mathcal{PT}$-symmetry breaking upon collision of their eigenvalues. This extends rich possibilities for control of localization of wave fields, soliton excitation thresholds, and propagation dynamics predicted in chirped conservative lattices [30-33] to the case of $\mathcal{PT}$-symmetric system. Moreover, we are mostly interested in evolution of the system *above* the symmetry-breaking threshold, where inclusion of dissipative nonlinearity enables formation of stable attractors, rather than in evolution in unbroken symmetry regime [34]. We discuss how properties of such dissipative solitons depend on the gain/loss amplitude.

We consider propagation of a light beam along the $z$-axis of the $\mathcal{PT}$-symmetric waveguide array created by simultaneous modulation of the refractive index and of gain/loss in the transverse $x$-direction. The dynamics of propagation is governed by the nonlinear Schrödinger equation for the dimensionless light field amplitude $\Psi$:

$$i\frac{\partial \Psi}{\partial z} = -\frac{1}{2}\frac{\partial^2 \Psi}{\partial x^2} - \mathcal{R}(x)\Psi - (1+i\alpha)|\Psi|^2\Psi. \qquad (1)$$

Here the transverse $x$ and longitudinal $z$ coordinates are normalized to the characteristic transverse scale $x_0$ and the diffraction length $L_{\text{dif}} = kx_0^2$, respectively; $k = 2\pi n_0/\lambda$ is the wavenumber; $n_0$ is the background refractive index; the last term describes focusing cubic nonlinearity and weak two-photon absorption with strength $\alpha$. The complex function $\mathcal{R} = \mathcal{R}_{\text{re}} - i\mathcal{R}_{\text{im}}$ describes shallow transverse modulation of the refractive index $\mathcal{R}_{\text{re}}(x) = p_{\text{re}} \sum_{k=-\text{K}}^{\text{K}} \mathcal{Q}(x - x_k)$ that is symmetric with respect to $x = 0$ and antisymmetric gain/loss profile $\mathcal{R}_{\text{im}}(x) = p_{\text{im}} \sum_{k=-\text{K}}^{\text{K}} (x - x_k) \mathcal{Q}(x - x_k)$, so that complex potential in (1) satisfies the $\mathcal{PT}$-symmetry condition $\mathcal{R}(x) = \mathcal{R}^*(-x)$. Here $p_{\text{re}}$ and $p_{\text{im}}$ characterize the actual depth of the refractive index modulation and gain/loss amplitude, respectively; $2\text{K}+1$ is the total number of waveguides in the array (below we consider the representative case of large array with $\text{K} = 20$); $\mathcal{Q}(x) = \exp(-x^6/d^6)$ describes super-Gaussian profiles of individual waveguides of width $d$ (see insets in Figs. 2 and 4 showing refractive index profile of such an array). Main distinctive feature of our arrays is that separation between neighboring waveguides in them changes in the transverse plane leading to larger density of waveguides in the center of the array or on its edge. The recursive formulas below describe how coordinates of waveguides vary in the region $x > 0$ (in the region $x < 0$ one has $x_{-k} = -x_k$):

$$x_{k+1} = x_k + 2\pi/[\Omega_{\min} + (k/\text{K})(\Omega_{\max} - \Omega_{\min})],$$
$$x_{k+1} = x_k + 2\pi/[\Omega_{\max} - (k/\text{K})(\Omega_{\max} - \Omega_{\min})], \quad (2)$$

where $x_0 = 0$, $\Omega_{\max}$ and $\Omega_{\min}$ stand for the maximal and minimal spatial frequencies, respectively. The first expression in (2) corresponds to the case when spacing between waveguides decreases (waveguide density increases) towards the edge of the array, while the second expression produces array with larger waveguide density in the center.

We start our analysis from consideration of linear modes of $\mathcal{PT}$-symmetric modulated array $\Psi = \psi(x)\exp(ibz)$, where $b = b_{\text{re}} + ib_{\text{im}}$ is the propagation constant that can be complex. Such modes can be obtained from Eq. (1), with omitted nonlinear terms. Further we use parameters $p_{\text{re}} = 8$, $d = 0.5$, $\Omega_{\max} = 1.3\pi$, $\Omega_{\min} = 0.65\pi$ at which single channel guides only one mode in the absence of gain/loss. Figure 1 illustrates variation of eigenvalues of modes of modulated waveguide array with increase of gain/loss amplitude $p_{\text{im}}$ for arrays with larger waveguide density in the center [panel (a)] or at the edge [panel (b)]. Shaded areas show two allowed bands that are separated by the gap. When $p_{\text{im}}$ grows the gap width gradually decreases. In both cases the $\mathcal{PT}$-symmetry gets broken when eigenvalues of the modes from the bottom of the first band (open black circles) collide with eigenvalues of the modes from the top of the second band (open red circles). The modes, whose collision leads to symmetry breaking are shown in Fig. 2 – usually these are modes with numbers $2\text{K}+1$ and $2\text{K}+2$, when modes are sorted such that largest $b_{\text{re}}$ value corresponds to mode with index 1 [see Figs. 2(a),(b) and 2(d),(e), corresponding to the regime of unbroken symmetry]. Mathematically, collision of eigenvalues is explained by the presence of corresponding nonzero off-diagonal elements $\langle \psi_{2\text{K}+1} | \mathcal{R}_{\text{im}} | \psi_{2\text{K}+2} \rangle$ in matrix representation of the operator $\mathcal{H} = \mathcal{H}_0 + i\mathcal{R}_{\text{im}}$, where imaginary part of the potential $\mathcal{R}_{\text{im}}$ is considered as a perturbation, and $\psi_k$ are the eigenmodes of unperturbed operator $\mathcal{H}_0 = -(1/2)\partial_x^2 - \mathcal{R}_{\text{re}}$. These off-diagonal elements, growing with $p_{\text{im}}$, cause collision of eigenvalues of corresponding matrix and their shift into the complex plane.

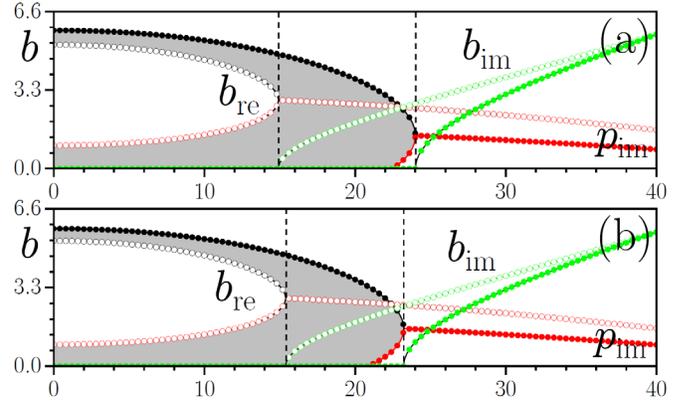

Figure 1. Real and imaginary parts of eigenvalues of representative modes of modulated arrays versus $p_{\text{im}}$. (a) Array with larger waveguide density in the center, (b) array with larger waveguide density at the edges. Shaded areas stand for allowed bands, where at least one mode has purely real eigenvalue. Real parts of eigenvalues that collide first are shown with open black and red circles, while corresponding imaginary part (only positive) is shown with green open circles. Real and imaginary parts of eigenvalues that collide last are depicted with solid circles.

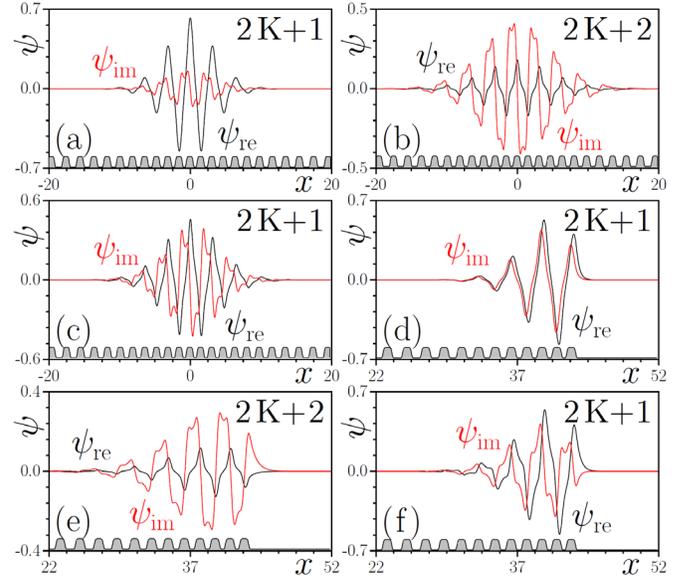

Figure 2. Examples of localized linear modes supported by modulated array at $p_{\text{im}} = 8$ (unbroken $\mathcal{PT}$-symmetry) (a),(b),(d),(e) and $p_{\text{im}} = 16$ (broken $\mathcal{PT}$-symmetry) (c),(f). Panels (a)-(c) correspond to array with larger waveguide density in the center, while (d)-(f) correspond to array with larger waveguide density at the edges. Panels (c),(f) show growing modes. Refractive index profile is shown as inset in the bottom.

Interestingly in the array with largest waveguide density in the center the $\mathcal{PT}$-symmetry occurs due to collision of eigenvalues of only two modes. However, in the array with largest density at the edges the symmetry is broken due to collision of two pairs of degenerate modes, i.e. after collision there appear two growing and two decaying modes with identical real part of the propagation constant $b_{\text{re}}$. The threshold value of $p_{\text{im}}$ for symmetry breaking is somewhat lower in the array with largest waveguide density in the center, see left dashed lines in Fig. 1. Symmetry breaking in our system always occurs due to collision of well-localized modes residing in the domains with largest densi-ty of waveguides. Thus, one can speak about "locally" broken $\mathcal{PT}$-symmetry because modes, whose field is concentrated in domains, where waveguides are sparser, still have real propagation constants if $p_{\text{im}}$ only slightly exceeds the threshold. The mode that first acquires complex propagation constant [see Figs. 2(c),(f), showing such modes in broken symmetry regime] will have highest growth rate (shown by the line with open green circles in Fig. 1) among all growing modes that appear with further increases of $p_{\text{im}}$ (see for example the line with solid circles in Fig. 1). This mode will dominate the dynamics in the broken $\mathcal{PT}$-symmetry regime. Adding nonlinear absorption $\alpha \neq 0$ allows to arrest the unlimited growth of amplitude of this mode and leads to formation of stable attractors either at the edge or in the center of waveguide array, depending on the type of modulation.

Such attractors were found numerically by solving the equation

$$b\psi = \frac{1}{2}\frac{d^2\psi}{dx^2} + \mathcal{R}\psi + (1+i\alpha)|\psi|^2\psi, \quad (3)$$

with an additional condition $\int[(\mathcal{R}-\mathcal{R}^*)|\psi|^2 + 2i\alpha|\psi|^4]dx = 0$ of power balance within stationary states allowing to determine the propagation constant that is not independent parameter in dissipative solitons. Inclusion of two-photon absorption $\alpha$ breaks $\mathcal{PT}$-symmetry of evolution Eq. (1). It substantially modifies gain-loss balance in the system and leads to new nonlinear stationary states. The dependences of power $U = \int|\psi|^2 dx$, peak amplitude $a = \max|\psi(x)|$, and integral width $w = U^2/\int|\psi|^4 dx$ of bulk and edge solitons on the amplitude of gain/loss $p_{\text{im}}$ are shown in Fig. 3. Dissipative solitons forming in the array with larger waveguide density in the center feature lower cutoff on $p_{\text{im}}$ (see left curves). The power of all types of solitons rapidly increases with increase of $p_{\text{im}}$, while initially fast growth of peak amplitude gradually saturates [Fig. 3(a)]. In contrast to usual conservative solitons, the width of dissipative states in this system increases with increase of their power [Fig. 3(b)].

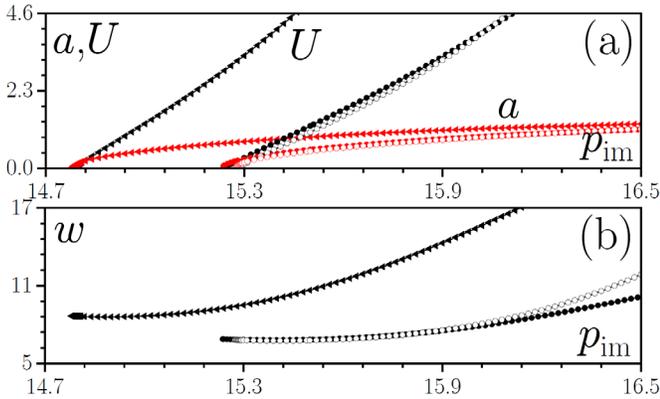

Figure 3. Peak amplitude $a$, power $U$ (a) and width $w$ (b) of dissipative solitons forming in the center of properly modulated array (triangles) and at its right (open circles) and left (solid circles) edges versus $p_{\text{im}}$ at $\alpha = 0.2$.

A noteworthy feature is that parameters of the soliton appearing on the left edge of the array are slightly different from those of soliton on the right edge. This happens because on the right the array terminates with an amplifying domain, while on the left it terminates with an absorbing domain in accordance with $\mathcal{PT}$-symmetric $\mathcal{R}(x)$ profile.

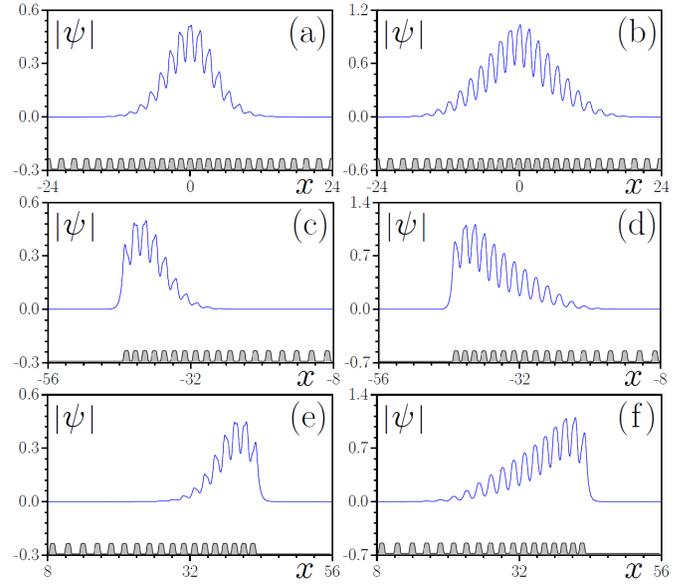

Figure 4. Dissipative solitons in the array with larger waveguide density in the center at (a) $p_{\text{im}} = 15$, $b \approx 2.84$ and (b) $p_{\text{im}} = 15.7$, $b \approx 3.47$. Solitons in the array with larger waveguide density at the edges at (c),(e) $p_{\text{im}} = 15.45$, $b \approx 2.77$, (d),(f) $p_{\text{im}} = 16.4$, $b \approx 3.54$ and $b \approx 3.64$, respectively. In all cases $\alpha = 0.2$.

The profiles of dissipative solitons emerging in the array with larger waveguide density in the center and at its edges are compared in Fig. 4. The asymmetry in the field modulus distribution within individual waveguides is obvious – on the one hand it is a consequence of broken $\mathcal{PT}$-symmetry (growing mode resides mostly in amplifying domains) and on the other hand it is due to modification of currents by nonlinear absorption, which by itself may cause asymmetries. The reason of asymmetry is thus different from that in non-$\mathcal{PT}$-symmetric complex potentials, where currents are asymmetric even in linear non-growing modes (and also in solitons) due to asymmetric shape of the potential [35-40]. Close to the threshold value of $p_{\text{im}}$, where the family of dissipative solitons emerges, their field modulus distribution $|\psi|$ resemble that of linear modes. Increasing gain/loss amplitude causes soliton expansion: even though the tails are still exponential, the soliton may develop a triangular shape clearly visible in panels (b), (d), and (f). At even larger $p_{\text{im}}$ values, the width of soliton becomes comparable with that of the waveguide array, however soliton in this regime is already dynamically unstable. Stability is encountered in the domain adjacent to the cutoff for existence of dissipative solitons on $p_{\text{im}}$. Interestingly, localization of the edge and central dissipative soliton could be remarkably improved by increasing the modulation rate simultaneously with the $p_{\text{re}}$ parameter responsible for the localization of the field inside waveguides.

As mentioned above, dissipative solitons that we consider here are stable attractors, therefore they can be excited from noise. Figures 5(a) and 5(b) illustrate the excitation process for array with larger waveguide density in the center and at the edge, respectively. In both cases the soliton forms after relatively short propagation distance, provided that $p_{\text{im}}$ value is not very far from the threshold. Importantly, solitons in the center of the array and on its edge are very robust objects if gain/loss amplitude $p_{\text{im}}$ is smaller than the value at which collision of second pair of linear eigenvalues occurs within gray areas between dashed lines in Figs. 1(a) and 1(b). Thus, in the array with larger waveguide density in the center, the first pair of linear modes collides at $p_{\text{im}} \approx 14.90$ (symmetry-breaking threshold), while the second pair of modes collides at $p_{\text{im}} \approx 15.75$ - in this interval nonlinear modes are stable. Stable propagation of central and edge dissipative solitons perturbed by small-amplitude noise up to large distances is shown in Figs. 5(c) and 5(d), respectively. If $p_{\text{im}}$ parameter exceeds the threshold value considerably, one observes spontaneous excitation of multiple modes that exhibit beatings leading to chaotic patterns, covering the entire waveguide array.

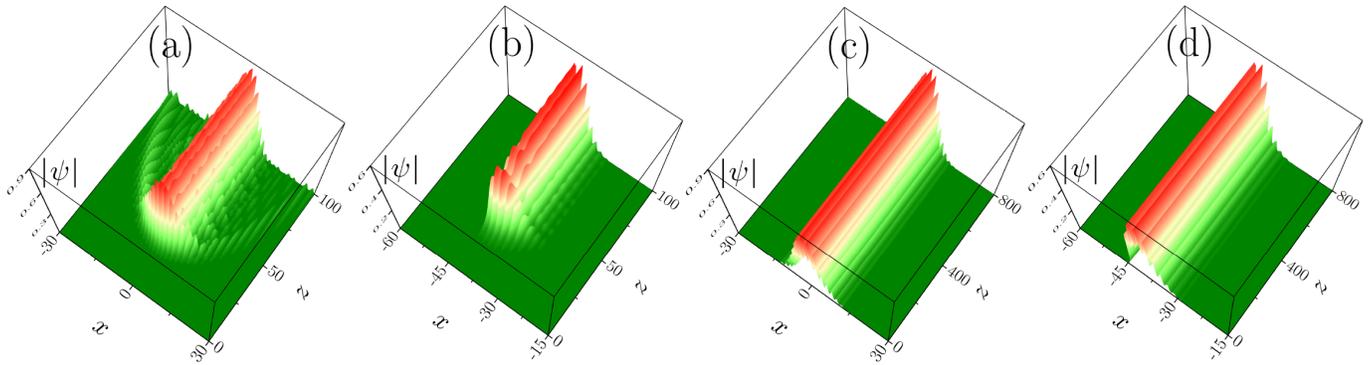

Figure 5. Excitation from small noise (a), (b) and stable propagation (c), (d) of dissipative solitons in the array with larger waveguide density in the center at $p_{\text{im}} = 15.4$ (a), (c) and array with larger waveguide density at the border at $p_{\text{im}} = 15.5$ (b), (d). In all cases $\alpha = 0.2$. In (c) $b \approx 3.23$, in (d) $b \approx 2.82$.

Summarizing, we found that in the modulated $\mathcal{PT}$-symmetric waveguide arrays, symmetry breaking may occur via collision of eigenvalue of well-localized edge or bulk modes concentrated in the domains where waveguides have smallest separation. In the regime with broken symmetry such modes give rise to dissipative solitons, whose location depends on the type of array modulation, provided that gain is compensated by the nonlinear absorption.

## References

1. C. M. Bender, Rep. Prog. Phys. **70**, 947 (2007).
2. S. V. Suchkov, A. A. Sukhorukov, J. H. Huang, S. V. Dmitriev, C. Lee, and Y. S. Kivshar, Las. Photon. Rev. **10**, 177 (2016).
3. V. V. Konotop, J. Yang, and D. A. Zezyulin, Rev. Mod. Phys. **88**, 035002 (2016).
4. S. Longhi, Europhys. Lett. **120**, 64001 (2017).
5. R. El-Ganainy, K. G. Makris, D. N. Christodoulides, and Z. H. Musslimani, Opt. Lett. **32**, 2632 (2007).
6. A. Guo, G. J. Salamo, D. Duchense, R. Morandotti, M. Volatier-Ravat, V. Aimez, G. A. Siviloglou, and D. N. Christodoulides, Phys. Rev. Lett. **103**, 093902 (2009).
7. C. E. Rüter, K. G. Makris, R. El-Ganainy, D. N. Christodoulides, M. Segev, and D. Kip, Nat. Phys. **6**, 192 (2010).
8. A. A. Sukhorukov, Z. Xu, and Y. S. Kivshar, Phys. Rev. A **82**, 043818 (2010).
9. H. Ramezani, T. Kottos, R. El-Ganainy, and D. N. Christodoulides, Phys. Rev. A **82**, 043803 (2010).
10. Y. Kominis, T. Bountis, and S. Flach, Sci. Rep. **6**, 33699 (2016).
11. Z. H. Musslimani, K. G. Makris, R. El-Ganainy, and D. N. Christodoulides, Phys. Rev. Lett. **100**, 030402 (2008).
12. F. K. Abdullaev, Y. V. Kartashov, V. V. Konotop, and D. A. Zezyulin, Phys. Rev. A **83**, 041805 (2011).
13. R. Driben and B. A. Malomed, Opt. Lett. **36**, 4323 (2011).
14. N. V. Alexeeva, I. V. Barashenkov, A. A. Sukhorukov, and Y. S. Kivshar, Phys. Rev. A **85**, 063837 (2012).
15. Y. He, D. Mihalache, X. Zhu, L. Guo, and Y. V. Kartashov, Opt. Lett. **37**, 2526 (2012).
16. U. Al Khawaja, S. M. Al-Marzoug, H. Bahlouli, and Y. S. Kivshar, Phys. Rev. A **88**, 023830 (2013).
17. Y. V. Kartashov, V. A. Vysloukh, and L. Torner, Opt. Lett. **41**, 4348 (2016).
18. Y. V. Kartashov, V. A. Vysloukh, V. V. Konotop, and L. Torner, Phys. Rev. A **93**, 013841 (2016).
19. T. Goldzak, A. A. Mailybaev, N. Moiseyev, Phys. Rev. Lett. **120**, 013901 (2018).
20. Y. V. Kartashov, C. Hang, V. V. Konotop, V. A. Vysloukh, G. Huang, and L. Torner, Las. Photon. Rev. **10**, 100 (2016).
21. S. Weimann, M. Kremer, Y. Plotnik, Y. Lumer, S. Nolte, K. G. Makris, M. Segev, M. C. Rechtsman, A. Szameit, Nat. Mater. **16**, 433 (2017).
22. I. V. Barashenkov, L. Baker, and N. V. Alexeeva, Phys. Rev. A **87**, 033819 (2013).
23. D. A. Zezyulin and V. V. Konotop, Phys. Rev. Lett. **108**, 213906 (2012).
24. A. Mock, Phys. Rev. A **93**, 063812 (2016).
25. C. Huang, F. Ye, and X. Chen, Phys. Rev. A **90**, 043833 (2014).
26. W. Walasik and N. M. Litchinitser, Sci. Rep. **6**, 19826 (2015).
27. Y. V. Kartashov, B. A. Malomed, and L. Torner, Opt. Lett. **39**, 5641 (2014).
28. J. D'Ambroise, P. G. Kevrekidis, and B. A. Malomed, Phys. Rev. E **91**, 033207 (2015).
29. T. Mayteevarunyoo, B. A. Malomed, and A. Reoksabutr, Phys. Rev E **88**, 022919 (2013).
30. Y. V. Kartashov, L. Torner, and V. A. Vysloukh, Opt. Express **13**, 4244 (2005).
31. Y. V. Kartashov, V. A. Vysloukh, and L. Torner, Phys. Rev. A **76**, 013831 (2007).
32. A. Szameit, Y. V. Kartashov, F. Dreisow, M. Heinrich, T. Pertsch, S. Nolte, A. Tünnermann, V. A. Vysloukh, and L. Torner, Opt. Lett. **33**, 1132 (2008).
33. V. A. Brazhnyi, V. V. Konotop, and V. Kuzmiak, Phys. Rev. A **70**, 043604 (2004).
34. L.-L. Gu, D.-C. Guo, L.-W. Dong, Opt. Express **23**, 12434 (2015).
35. E. N. Tsoy, I. M. Allayarov, and F. K. Abdullaev, Opt. Lett. **39**, 4215 (2014).
36. V. V. Konotop and D. A. Zezyulin, Opt. Lett. **39**, 5535 (2014).
37. J. Yang, Stud. Appl. Math. **132**, 332 (2014).
38. Y. Kominis, Phys. Rev. A **92**, 063849 (2015).
39. Y. Kominis, Opt. Commun. **334**, 265 (2015).
40. S. D. Nixon and J. Yang, Stud. Appl. Math. **136**, 459 (2016).